\documentclass[12pt]{article}

\usepackage{epsfig,graphics}

\begin{document}

\title{\bf The random link approximation for the Euclidean
traveling salesman problem}
\author{N.J.~Cerf\thanks{Present address: W.K.~Kellogg Radiation
                         Laboratory, California Institute of
                         Technology, Pasadena, CA 91125, USA}\ ,
        J.~Boutet de Monvel, O.~Bohigas,\\ 
        O.C.~Martin\thanks{E-mail address: martino@ipno.in2p3.fr}\ \ and
	A.G.~Percus\thanks{E-mail address: percus@ipno.in2p3.fr}\\
\\
{\normalsize Division de Physique Th\'eorique\thanks{Unit\'e de Recherche
des Universit\'es Paris XI et Paris VI associ\'ee au C.N.R.S.}} \\
{\normalsize Institut de Physique Nucl\'eaire, Universit\'e\ Paris-Sud} \\
{\normalsize F--91406 Orsay Cedex, France} \\ }

\date{}
\maketitle
\begin{abstract}
\medskip
The traveling salesman problem (TSP) consists of finding the
length of the shortest closed tour visiting $N$ ``cities.''  We
consider the 
Euclidean TSP where the cities are distributed
randomly and independently in a $d$-dimensional
unit hypercube.
Working with periodic boundary conditions and inspired by
a remarkable universality in the $k$th nearest neighbor
distribution, we find for the average optimum tour length
$\langle L_E\rangle = \beta_E(d)\,N^{1-1/d}\,[1+O(1/N)]$ with
$\beta_E(2)=0.7120 \pm 0.0002$ and 
$\beta_E(3)=0.6979 \pm 0.0002$. 
We then derive analytical predictions for these quantities
using the random link approximation, where the lengths
between cities are taken as independent random variables. 
From the ``cavity'' equations developed by Krauth, M\'ezard
and Parisi, we calculate the associated random link
values $\beta_{RL}(d)$.
For $d=1,2,3$, numerical results show that the random link approximation
is a good one, with a discrepancy of less than 2.1\% between 
$\beta_E(d)$ and $\beta_{RL}(d)$.  For large $d$, we argue that the
approximation is exact up to $O(1/d^2)$ and give a conjecture for
$\beta_E(d)$, in terms of a power series in $1/d$, specifying both
leading and subleading coefficients.

\end{abstract}

\bigskip
\noindent {PACS numbers: 02.60.Pn, 02.70.Lq, 64.60.Cn}
\par
\vspace{1cm}
\centerline{Appeared in {\it Journal de Physique I\/} {\bf 7}, 117--136,
January 1997}

\bigskip

\newpage
\baselineskip=20pt
\section{Introduction}
\label{sect_introduction}
Given $N$ ``cities'' and the distances between them, the
traveling salesman problem (TSP) consists of finding the
length of the shortest closed ``tour'' (path) visiting every
city exactly once, where the tour length is the sum of the
city-to-city distances along the tour.
The TSP is NP-complete, which suggests that there is
no general algorithm capable of finding the optimum tour in
an amount of time polynomial in $N$.  The problem is thus simple to
state, but very difficult to solve.  It also happens to be
the most well known combinatorial optimization problem,
and has attracted interest from a wide
range of fields.  In operations research, mathematics and
computer science, researchers have concentrated on algorithmic aspects.
A particular focus has been on heuristic algorithms --- algorithms
which do not guarantee optimal tours --- for cases where exact
methods are too slow to be of use.
The most effective heuristics are based on local search methods,
which start with a non-optimal tour and iteratively improve the
tour within a well-defined ``neighborhood''; a famous example
is the Lin-Kernighan heuristic \cite{LinKernighan}.  More recent
efforts have involved combining local search and non-deterministic
methods, in order to refine heuristics to the point where they
give good enough solutions for practical purposes; a powerful
such technique is Chained Local Optimization \cite{MartinOtto_AOR}.

Over the last fifteen years, physicists have increasingly been
drawn to the TSP as well, and particularly to {\it stochastic\/}
versions of the problem, where instances are randomly chosen from
an ensemble.  The motivation has often been to find properties
applicable to a large class of disordered systems, either through
good approximate methods or through exact analytical approaches.
In our work, we consider two such stochastic TSPs.  The first,
the Euclidean TSP,
is the more classic form of the problem: $N$ cities are placed
randomly and independently in a $d$-dimensional
hypercube, and the distances between cities are defined by the
Euclidean metric.  The second, the random link TSP, is a 
related problem
developed within the context of disordered systems: rather
than specifying the positions of cities, we specify the lengths
$l_{ij}$ separating cities $i$ and
$j$, where the $l_{ij}$ are taken to be independent, identically
distributed random variables.
The appeal of the random link problem is, on the one hand,
that an analytical approach exists for solving it
\cite{MezardParisi_86b,KrauthMezard}, and on the
other hand, that when certain correlations are neglected this
TSP can be made to resemble the Euclidean TSP.  We therefore
consider the random link problem as a
{\it random link approximation\/} to the (random point) Euclidean
problem.  Researchers outside of physics remain largely unaware of
the analytical progress made on the random link TSP; one of our
hopes is to demonstrate how these results are of direct interest
in problems where the aim is to find the optimum Euclidean
TSP tour length.

Our approach in this paper is then to examine both the Euclidean
problem and the random link problem --- the latter for its own
theoretical interest as well as for a better understanding of the
Euclidean case.  We begin by considering in depth the Euclidean TSP,
including a review of previous work.  We find that, given periodic
boundary conditions (toroidal geometry), the Euclidean optimum tour
length $L_E$ averaged
over the ensemble of all possible instances has the finite size
scaling behavior
\begin{equation}
\langle L_E\rangle = \beta_E(d)\,N^{1-1/d}\,\left[ 1+O\left(
\frac{1}{N}\right)\right]\mbox{.}
\end{equation}
From simulations, we extract very precise numerical values
for $\beta_E(d)$ at $d=2$ and $d=3$;
methodological and numerical procedures are detailed in the appendices.
We also give numerical evidence that the
probability distribution of $L_E$ becomes Gaussian in the large $N$ limit. 
In addition to these TSP results, we find a surprising
universality in the scaling of the mean distance between $k$th
nearest neighbors, for points randomly distributed in the
$d$-dimensional hypercube.
Finally, we discuss the expected behavior of $\beta_E(d)$ in the
large $d$ limit.

In the second part of the paper we discuss the random link problem,
considering it as an approximation to the Euclidean problem.
Making use of the cavity method, we compare the random link
$\beta_{RL}(d)$ with the Euclidean $\beta_E(d)$ values obtained from our
simulations.  We find that the random link approximation is correct to
within 2\% at $d=2$ and $3$.  The rest of the section
studies the large $d$ limit of the random link model and its implications
for the Euclidean TSP.  We examine analytically how
$\beta_{RL}(d)$ scales at large $d$, and we relate the $1/d$
coefficient of the associated power series to an underlying
$d$-independent ``renormalized'' model.
Finally, we present a theoretical analysis based on 
the Lin-Kernighan heuristic, suggesting strongly that the relative
difference between $\beta_{RL}(d)$ and $\beta_E(d)$ is positive
and of $O(1/d^2)$.  The random link results then lead to our large $d$
Euclidean conjecture:
\begin{equation}
\label{eq_strong_conjecture}
\beta_E(d) = \sqrt{\frac{d}{2\pi e}}\,(\pi d)^{1/2d}
\left[ 1+\frac{2-\ln 2-2\gamma}{d}
+ O\left(\frac{1}{d^2}\right)\right]\mbox{,}
\end{equation}
where $\gamma$ is Euler's constant.

\newpage

\section{The Euclidean TSP}
\label{sect_E}

\subsection{Scaling at large N}
\label{sect_E_Scaling}

\begin{figure}[t]
\begin{center}
\begin{picture}(289,268)
\epsfig{file=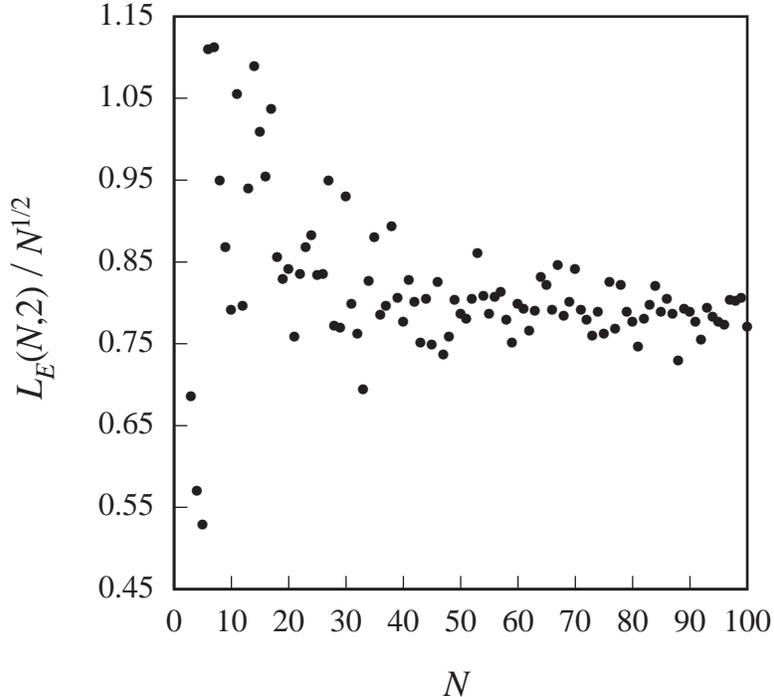}
\end{picture}
\caption{\small Self-averaging of 2-D Euclidean TSP optimum:
convergence of $L_E(N,2)/N^{1/2}$ on a sequence
of random instances at increasing $N$.}
\label{fig_selfav}
\end{center}
\end{figure}

One of the earliest analytical results for the Euclidean TSP
is due to Beardwood, Halton and Hammersley \cite{BHH} (BHH).
The authors considered $N$ cities, distributed randomly and independently
in a $d$-dimensional volume with distances between cities given by the
Euclidean metric.  They showed that, when the volume is the unit
hypercube and the distribution of cities uniform, $L_E/N^{1-1/d}$ is
self-averaging.  This means that with probability 1,
\begin{equation}
\label{eq_bhh}
\lim_{N\to\infty} \frac{L_E}{N^{1-1/d}} = \beta_E(d)\mbox{,}
\end{equation}
where $\beta_E(d)$ is independent of the randomly chosen instances.
This property is illustrated in Figure \ref{fig_selfav}.
(In fact, the BHH result is more general than this and concerns
an arbitrary volume and arbitrary form of the density of cities.)
For a physics audience this large $N$ limit is equivalent,
in appropriate units, to an infinite volume limit at constant
density.  $L_E/N^{1-1/d}$ then corresponds to an energy density
that is self-averaging and has a well-defined infinite volume
limit.
The original proof by BHH is quite
complicated; simpler proofs have since been given by Karp and 
Steele \cite{KarpSteele,Steele}.

One of our goals is to determine $\beta_E(d)$.  BHH gave rigorous
lower and upper bounds as a function of dimension.  For any given
instance, a trivial lower bound on $L_E$ is the sum over all cities
$i$ of the distance between $i$ and its nearest neighbor in space.
In fact, since a tour at best links a city with its {\it two\/} nearest
neighbors, this bound can be improved upon by summing, over all $i$,
the mean of $i$'s nearest and next-nearest neighbor distances.
Taking the ensemble average of this quantity (that is, the average
over all instances)
leads to the best analytical lower bound to date.
For upper bounds, BHH introduced a heuristic algorithm, now known
as ``strip,'' in order to generate near-optimal tours (discussed also in
a paper by Armour and Wheeler \cite{ArmourWheeler}).
In two dimensions the method involves dividing the square into 
adjacent columns or strips, and sequentially visiting the cities on
a given strip according to their positions along it. 
The respective lower and upper bounds give
$0.6250 \le \beta_E(2) \le 0.9204$.

In addition to bounds, it is possible to obtain numerical
{\it estimates\/} for $\beta_E(d)$.
BHH used two instances, $N=202$ and $N=400$, from which they
estimated $\beta_E(2) \approx 0.749$ using hand-drawn tours.
Surprisingly little has been done to improve upon this value
in two dimensions, and essentially nothing in higher dimensions.
Stein \cite{Stein_Thesis} has found $\beta_E(2) \approx 0.765$,
which is frequently cited.  Only recently have better values been
obtained, but as they come from near-optimal tours found by heuristic
algorithms, they should be
considered more as upper bounds than as estimates. 
Using a local search heuristic known as ``3-opt'' \cite{Lin},
Ong and Huang \cite{OngHuang} have found $\beta_E(2) \le 0.743$;
using another heuristic, ``tabu'' search, Fiechter \cite{Fiechter}
has found $\beta_E(2) \le 0.731$;
and using a variant of simulated annealing, Lee and 
Choi \cite{LeeChoi} have found $\beta_E(2) \le 0.721$.  In what follows
we shall show what is needed for a more precise estimate of
$\beta_E(d)$ with, furthermore, a way to quantify the associated
error.

\subsection{Extracting $\beta_E(d)$}
\label{sect_E_Methodology}

As $N \to \infty$, $L_E / N^{1-1/d}$ converges with
probability 1 to the instance-independent $\beta_E(d)$.
Our estimate of $\beta_E(d)$ must rest on some assumptions, though,
since only finite values of $N$ are accessible numerically.
Note first that at values of $N$ where computation times are reasonable,
$L_E$ has substantial instance-to-instance fluctuations.
To reduce and at the same time quantify these fluctuations, we
average over a large number of instances.  We thus consider the
numerical mean of $L_E$ over the instances sampled, which itself
satisfies the asymptotic relation (\ref{eq_bhh}) but with a smoother
convergence.  To extract $\beta_E(d)$, we must understand precisely
what this convergence in $N$ is.

If cities were randomly distributed in the hypercube with open
boundary conditions, the cities near the boundaries would have fewer
neighbors and therefore lengthen the tour.  In standard statistical
mechanical systems at constant density, boundary effects lead to
corrections of the form surface over volume.  For the TSP at constant
density, the volume grows as $N$ and the surface as $N^{1-1/d}$.
In a $d$-dimensional unit hypercube, then, the ensemble average of $L_E$
would presumably have the large $N$ behavior
\begin{equation}
N^{1-1/d}\,\beta_E(d)\left( 1 + \frac{A}{N^{1/d}} + \frac{B}{N^{2/d}} +
\cdots\right)\mbox{.}
\end{equation}
In order to extract $\beta_E(d)$ numerically, it would be
necessary to perform a fit which includes these corrections.
A reliable numerical fit, however, must have few adjustable
parameters, and the slow convergence of this series would
prevent us from extracting $\beta_E(d)$ to high accuracy.
We therefore have chosen to eliminate these boundary (surface)
effects by using {\it periodic\/} boundary conditions in all
directions.  This should not change $\beta_E(d)$, but leaves us
with fewer adjustable parameters and a faster
convergence, enabling us to work with smaller
values of $N$ where numerical simulations are not too slow.

For the hypercube with periodic boundary conditions, let us
introduce the notation
\begin{equation}
\beta_E(N,d)\equiv\frac{\langle L_E(N,d)\rangle}{N^{1-1/d}}\mbox{,}
\end{equation}
where $\langle L_E\rangle$ is the average of $L_E$ over the
ensemble of instances.  ($\beta_E(N,d)$ is, in
physical units, the zero-temperature energy density.)
We then wish to understand how $\beta_E(N,d)$
converges to its large $N$ limit, $\beta_E(d)$.
In standard statistical mechanical systems,
there is a characteristic correlation length $\xi$.
Away from a critical point, $\xi$ is finite,
and finite size corrections decrease as $e^{-W/\xi}$,
where $W$ is a measure of the system ``width.''
At a critical point, $\xi$ is infinite, and finite size
corrections decrease as a power of $1/W$.
For {\it disordered\/} statistical systems, however, this
picture must be modified.
Even if $\xi$ is finite for each instance in the
ensemble, the fluctuating disorder can still give rise to
power-law corrections for ensemble averaged quantities.
In the case of the TSP, this is particularly clear: the
disorder in the positions of the cities induces large
finite size effects even on simple geometric quantities.

\begin{figure}[!b]
\begin{center}
\vspace{20pt}
\begin{picture}(348,222)
\epsfig{file=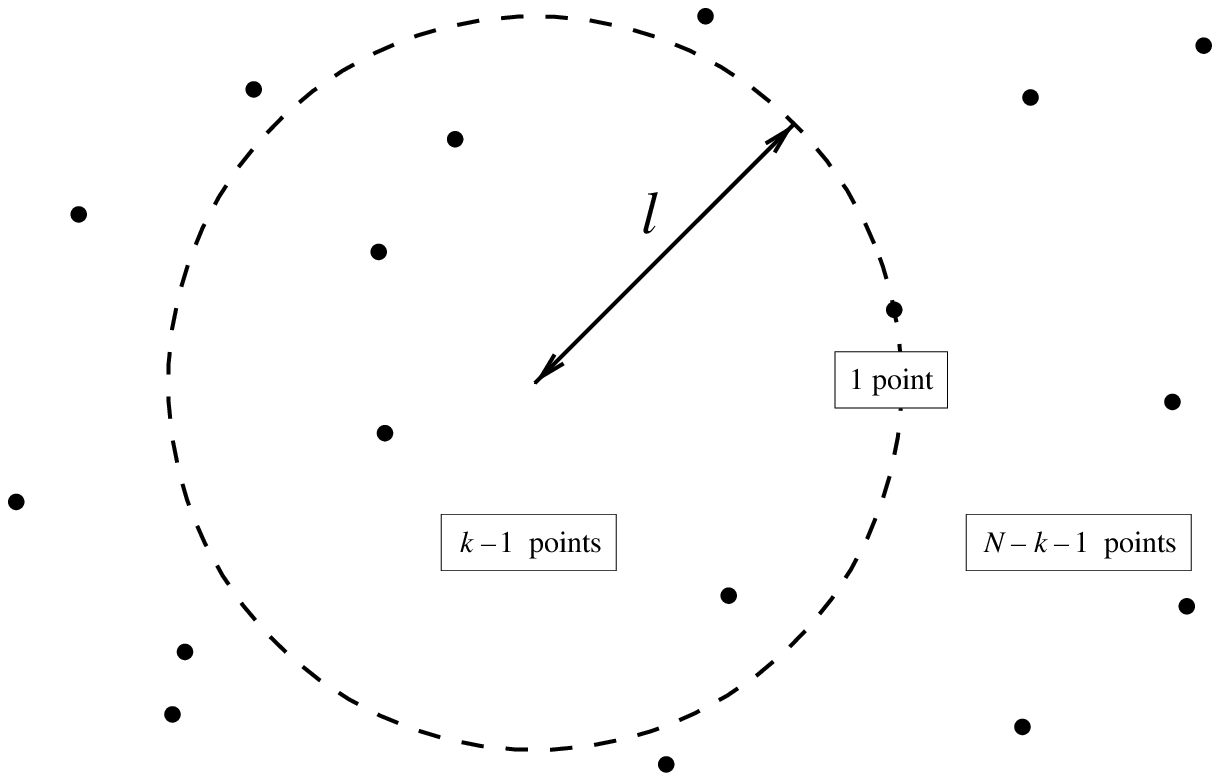}
\end{picture}
\caption{\small A point's $N-1$ neighbors: $k-1$ nearest neighbors are
within distance $l$, $k$th nearest neighbor is at $l$, and remaining
$N-k-1$ points are beyond $l$.}
\label{fig_nn}
\vspace{-20pt}
\end{center}
\end{figure}

To see how this might affect the convergence of $\beta_E(N,d)$,
consider the following.
For a given configuration of $N$ points, call $D_k(N,d)$ the distance
between a point and its $k$th nearest neighbor, where $k=1,\dots,N-1$.
Take the points to be distributed randomly and uniformly in the
unit hypercube.  Let us find $\langle D_k(N,d)\rangle$.  Under
periodic boundary conditions,
the probability density $\rho(l)$ of finding a point
at distance $l$ from another point is simply equal (for $0\le l\le 1/2$)
to the surface area at radius $l$ of the $d$-dimensional sphere:
\begin{equation}
\rho(l) = \frac{d\,\pi^{d/2}}{\Gamma(d/2+1)}\,l^{d-1}\mbox{.}
\label{eq_probdens}
\end{equation}
The probability of finding a point's $k$th nearest neighbor at distance
$l$ (see Figure \ref{fig_nn}) is equal to the probability of finding
$k-1$ (out of $N-1$) points within $l$, one point at $l$ and the
remaining $N-k-1$ points beyond $l$:
\begin{eqnarray}
P[D_k(N,d)=l] &=& {N-1 \choose k-1}\,\left[ \int_0^l \rho(l')\,
dl'\right]^{k-1} \times\, (N-k)\,\rho(l)\,\times\,
\left[1-\int_0^l \rho(l')\,dl'\right]^{N-k-1}\\
&=& {N-1 \choose k-1}\, (N-k)\, d\left[ \frac{\pi^{d/2}}
{\Gamma(d/2+1)}\right]^k l^{dk-1}
\left[ 1 - \frac{\pi^{d/2}}{\Gamma(d/2+1)} l^d
\right]^{N-k-1}\mbox{,}
\end{eqnarray}
giving the ensemble average
\begin{equation}
\langle D_k(N,d)\rangle = {N-1 \choose k-1}\,(N-k)\, d\left[ \frac{\pi^{d/2}}
{\Gamma(d/2+1)}\right]^k
\int_0^{1/2} l^{dk}
\left[ 1 - \frac{\pi^{d/2}}{\Gamma(d/2+1)} l^d \right]^{N-k-1}
dl\,+\,\cdots
\label{lbeqn}
\end{equation}
where the corrections are due to the $l>1/2$ case, and are
exponentially small in $N$.

Recognizing the integral, up to a simple change of variable, as a
Beta function ($B(a,b)\equiv\int_0^1 t^{a-1}(1-t)^{b-1}\,dt =
\Gamma(a)\Gamma(b)/\Gamma(a+b)$) plus a further remainder term
exponentially small in $N$, we see that
\begin{eqnarray}
\langle D_k(N,d)\rangle &=&\frac{\Gamma(d/2+1)^{1/d}}{\sqrt{\pi}}\,
\frac{\Gamma(k+1/d)}{\Gamma(k)}\,
\frac{\Gamma(N)}{\Gamma(N+1/d)}\,+\,\cdots\\
&=&\frac{\Gamma(d/2+1)^{1/d}}{\sqrt{\pi}}\,
\frac{\Gamma(k+1/d)}{\Gamma(k)}\,
N^{-1/d}\left[ 1 + \frac{1/d - 1/d^2}{2N} +
O\left( \frac{1}{N^2}\right) \right]\mbox{.}
\label{eq_nn}
\end{eqnarray}

We are confronted here with a remarkable, and hitherto unexplored,
universality:
the {\it exact\/} same $1/N$ series gives
the $N$-dependence regardless of $k$.  The same finite
size scaling behavior therefore applies to all $k$th nearest
neighbor distances.  

It might be hoped then that the typical link length in optimum
tours would have this $N$-dependence, and that $\beta_E(N,d)$
would therefore have the same $1/N$ expansion.  This is not
quite the case.  The link between cities $i$
and $j$ figures in the average $\langle D_k(N,d)\rangle$ whenever
$j$ is the $k$th neighbor of $i$; it figures in
$\beta_E(N,d)$, however, only when it belongs to the optimal
tour.  Two different kinds of averages are being taken, and so
finite size corrections need not be identical.
Nevertheless, it remains
plausible that $\beta_E(N,d)$ has a $1/N$ series expansion,
albeit a different one from (\ref{eq_nn}).  While we cannot prove
this property, it is confirmed by an analysis of our numerical data.

Our approach to finding $\beta_E(d)$ is thus as follows:
(i) we consider the ensemble
average $\langle L_E\rangle$, rather than $L_E$ for a given instance,
in order to have a quantity with a well-defined dependence on $N$;
(ii) we use periodic boundary conditions to eliminate 
surface effects; (iii) we sample the ensemble using numerical
simulations, and measure $\beta_E(N,d)$ within well controlled 
errors; (iv) we extract $\beta_E(d)$
by fitting these values to a $1/N$ series.

\subsection{Finite size scaling results}
\label{sect_E_Finite}

Let us consider the $d=2$ case in detail.  We found the most effective
numerical optimization methods for our purposes to be the local
search heuristics Lin-Kernighan (LK) \cite{LinKernighan} and
Chained Local Optimization (CLO) \cite{MartinOtto_AOR} mentioned
in the introduction.  Both heuristics,
by definition, give tour lengths that are not always optimal.
However, it is not necessary that the optimum be found 100\%
of the time: there is already a significant statistical error
arising from instance-to-instance fluctuations, and so
a further systematic error due to non-optimal tours
is acceptable as long as this error is kept negligible compared
to the statistical error.  Our methods, along with
relevant numerical details, are discussed in the appendices.
For the present purposes, let us simply mention the general nature
of the two heuristics used.  LK works by performing a
``variable-depth'' local search, as discussed further
in Section \ref{sect_RL_estimate}.  CLO works by an iterative
process combining LK optimizations with random
perturbations to the tour, in order to explore many different
local neighborhoods.
We used LK for ``small'' $N$ values ($N \le 17$), averaging
over 250,000 instances at each value of $N$, and we used CLO for
``large'' $N$ values ($N=30$ and $N=100$), averaging
over 10,000 and 6,000 instances respectively.

\begin{figure}[!b]
\begin{center}
\vspace{20pt}
\begin{picture}(300,274)
\epsfig{file=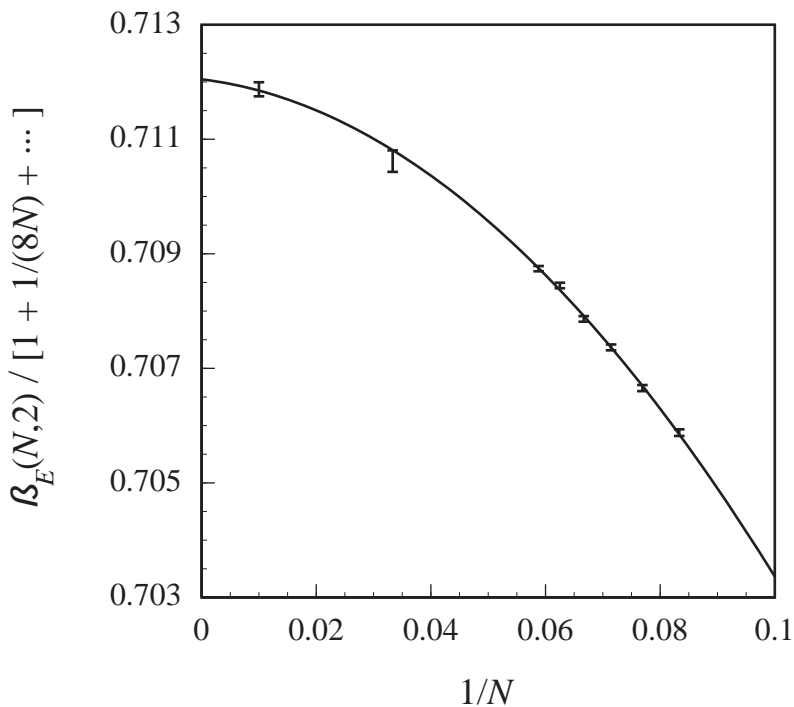}
\end{picture}
\caption{\small Finite size dependence of the rescaled 2-D Euclidean TSP
optimum.  Best fit ($\chi^2=5.56$) gives:
$\beta_E(N,2)/[1 + 1/(8N) + \cdots] = 0.7120
(1 - 0.0171/N - 1.048/N^2)$.  Error bars represent statistical errors.}
\label{fig_fss}
\vspace{-20pt}
\end{center}
\end{figure}

We fitted our resulting $\beta_E(N,d)$ estimates to a truncated
$1/N$ series: the fits are good, and are stable with respect to
the use of sub-samples of the data.
For a fit of the form $\beta_E(N,d) = \beta_E(d) (1 + A/N +
B/N^2)$, we find $\beta_E(2) = 0.7120 \pm 0.0002$, with
$\chi^2=5.57$ for 8 data points and 3 fit parameters (5 degrees
of freedom).  Our error estimate for $\beta_E(2)$ is obtained by the
standard method of performing fits using a range of fixed values
for this parameter: the error bar $\pm 0.0002$ is determined by
the values of $\beta_E(2)$ which make $\chi^2$ exceed its original
result by exactly 1, i.e., making $\chi^2=6.57$ in this case.

It is possible to extract another $\beta_E(N,d)$ estimate by
making direct use of the universality discussed previously:
the universal $1/N$ series in (\ref{eq_nn}) suggests that there
will be a faster convergence if we use the rescaled data
$\beta_E(N,2) / [1 + 1/(8N) +\cdots ]$.
This also has the appealing property of leading to a function
monotonic in $N$, as shown in Figure \ref{fig_fss}.  We find
\begin{equation}
\frac{\beta_E(N,2)}{1 + 1/(8N) + \cdots} \approx 0.7120\left(
1 - \frac{0.0171}{N} - \frac{1.048}{N^2}\right)
\end{equation}
with the leading term having the same error bar of $\pm 0.0002$ as before.
Note that the $1/N$ term in the fit is small ---
2 orders of magnitude smaller than the leading order coefficient ---
and so to first order the $1+1/8N+\cdots$ series
is itself a good approximation.

The same methodology was applied to the $d=3$ case.
The $\chi^2$s again confirmed the functional form of the
fit, and we find from our data $\beta_E(3) = 0.6979 \pm 0.0002$.
Also, since our initial work \cite{PercusMartin_PRL}, 
Johnson {\it et al.\/}\ have performed simulations at $d=2,3,4$,
obtaining results \cite{Johnson_HK} consistent with ours:
$\beta_E(2)\approx 0.7124$, $\beta_E(3)\approx 0.6980$ and
$\beta_E(4)\approx 0.7234$.

\subsection{Distribution of optimum tour lengths}
\label{sect_E_Distribution}

While BHH and others \cite{KarpSteele,Steele} have shown that
the variance of $L_E/N^{1-1/d}$ goes to zero as $N \to \infty$
(see also Figure \ref{fig_selfav}),
they have not determined how fast this variance decreases.
More generally, one
might ask how the {\it distribution\/} of $L_E/N^{1-1/d}$
behaves as $N \to \infty$.
We are aware of only one result, by Rhee and Talagrand
\cite{RheeTalagrand}, showing that the probability of finding
$L_E$ with $|L_E - \langle L_E\rangle | > t$
is smaller than $K \exp (-t^2 / K)$ for some $K$.
Unfortunately this is not strong enough to give bounds on the variance.

Let us characterize the distribution at $d=2$
by numerical simulation.  For motivation, consider the analogy
between $L_E/N^{1-1/d}$ and $E/V$, the energy
density in a disordered statistical system. 
If the system's correlation
length $\xi$ is finite (the system is not critical),
$E/V$ has a distribution which becomes Gaussian when $V \to \infty$.
This is because as the subvolumes increase, the energy 
densities in each subvolume become uncorrelated; the central limit
theorem then applies. A consequence is that $\sigma^2$, the 
variance of $E/V$, decreases as $V^{-1}$.
If $\xi$ is infinite (the system
is critical), then in general the distribution of $E/V$
is not Gaussian.  In both cases though, the self-averaging
of $E/V$ suggests that the scaling variable
$X = (E - \langle E \rangle )/ \sigma V$ has a limiting
distribution when $V \to \infty$. 

\begin{figure}[t]
\begin{center}
\begin{picture}(288,274)
\epsfig{file=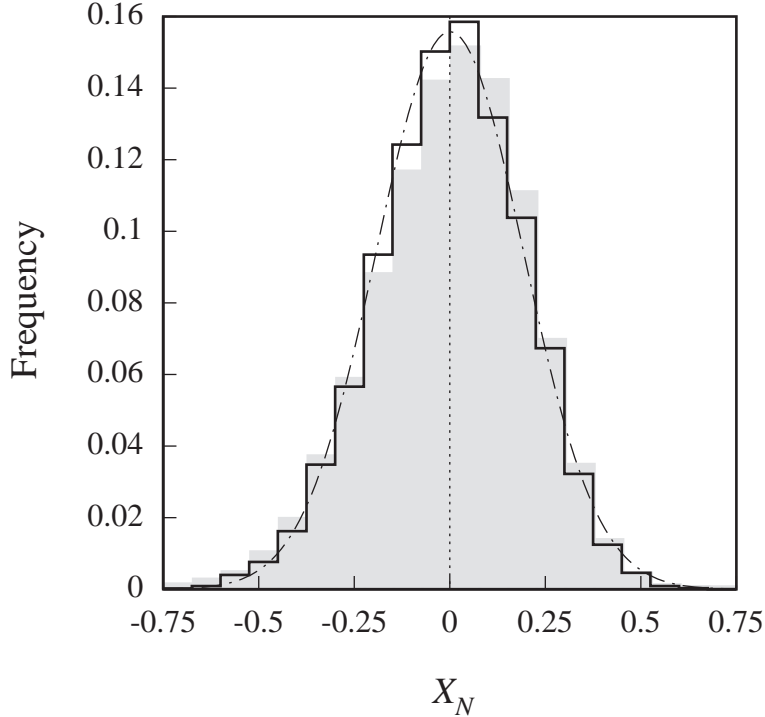}
\end{picture}
\caption{\small Distribution of 2-D Euclidean TSP scaling variable
$X_N = (L_E - \langle L_E\rangle ) / N^{1/2-1/d}$.  Shaded region
is for $N=12$ (100,000 instances used) and solid line is for $N=30$
(10,000 instances used).  Superimposed curve shows (extrapolated)
limiting Gaussian.}
\label{fig_dist}
\end{center}
\end{figure}

In the case of the TSP, it can be argued using a theoretical analysis
of the LK heuristic
that at $d\ge 2$ the system is not critical.
By analogy with $E/V$, if we take subvolumes to contain a fixed number
of cities, the central limit theorem then suggests that $L_E/N^{1-1/d}$
has a Gaussian distribution
with $\sigma^2$ decreasing as $N^{-1}$.  The scaling variable $X_N =
(L_E - \langle L_E\rangle ) / N^{1/2-1/d}$ should consequently
have a Gaussian distribution with a finite width for
$N \to \infty$ (and at $d\ge 2$).
Numerical results at $d=2$ (see Figure \ref{fig_dist}) give good
support for this.

\subsection{Conjectures on the large $d$ limit}
\label{sect_E_Large_d}

In most statistical mechanics problems, the large dimensional limit 
introduces simplifications because fluctuations become negligible.
For the TSP, can one expect $\beta_E(d)$ to have a simple limit
as $d \to \infty$?
Again, consider the property of the $k$th nearest
neighbor distance $D_k$.
In the large $N$ limit, (\ref{eq_nn}) gives
\begin{eqnarray}
N\langle D_k(N,d)\rangle&\sim&N^{1-1/d}\,
\frac{\Gamma(d/2+1)^{1/d}}{\sqrt{\pi}}
\,\frac{\Gamma(k+1/d)}{\Gamma(k)}\mbox{, or at large $d$,} \\
&\sim&N^{1-1/d}\sqrt{\frac{d}{2\pi e}}\,(\pi d)^{1/2d} \left[ 1 +
\frac{A_k}{d}+\cdots\right]\mbox{,}
\label{eq_largek}
\end{eqnarray}
where $A_k\equiv -\gamma+\frac{1}{k-1}+\frac{1}{k-2}+\cdots$
($\gamma$ is Euler's constant).  Notice that $A_k\sim\ln k$ at large
$k$.  This suggests strongly that unless the ``typical'' $k$ used
in the optimum tour grows exponentially in $d$, we may write
for $d\to\infty$:
\begin{equation}
\label{eq_dimscal}
\beta_E(d) = \lim_{N\to\infty}\frac{\langle L_E(N,d)\rangle}{N^{1-1/d}}
\sim \sqrt{\frac{d}{2\pi e}}\,(\pi d)^{1/2d}
\left[ 1 + O\left( \frac{1}{d}\right) \right]\mbox{.}
\end{equation}
Up to $O(1/d)$, this expression is identical to the BHH
lower bound on $\beta_E(d)$ discussed in Section \ref{sect_E_Scaling},
given by the large $N$ limit of $N^{1/d} \langle D_1(N,d)+D_2(N,d)\rangle /2$.

A weaker conjecture than (\ref{eq_dimscal}) has been proposed
by Bertsimas and van Ryzin \cite{BertsimasVanRyzin}:
\begin{equation}
\label{eq_Bertsimas}
\beta_E(d)\sim\sqrt{d/2\pi e}\mbox{ as }d\to\infty\mbox{.}
\end{equation}
This limiting behavior was motivated by an analogous result for a related
combinatorial optimization problem, the minimum spanning tree.
Unfortunately, there is no proof of either (\ref{eq_dimscal})
or (\ref{eq_Bertsimas}); in particular, the
upper bound on $\beta_E(d)$ given by strip, discussed in
Section \ref{sect_E_Scaling}, behaves 
as $\sqrt{d/6}$ at large $d$.  Thus if the conjectures are true, the 
strip construction leads asymptotically to tours which are on average
1.69 times too long.  Can we derive stronger upper bounds? 
A number of heuristic construction
methods should do better than strip, but there are
no reliable calculations to this effect.  The only improvements
over the BHH results are due to 
Smith \cite{Smith_Thesis}, who generalized the strip algorithm by
optimizing the shape of the strips, leading to an upper bound which
is $\sqrt{2}$ times greater than the predictions of (\ref{eq_dimscal})
and (\ref{eq_Bertsimas}) at large $d$.

In spite of our inability to derive an upper bound which,
together with the BHH lower bound,
would confirm the two conjectures for $d \to \infty$,
we are confident that
(\ref{eq_dimscal}) and (\ref{eq_Bertsimas}) are true because
of non-rigorous yet convincing arguments. 
One is a proof that (\ref{eq_Bertsimas})
is satisfied for the TSP if it is satisfied for another related
combinatorial optimization problem
(see \ref{app_heuristics} for details).
A more powerful argument,
presented in Section \ref{sect_RL_estimate}, relies on
a theoretical analysis of the LK heuristic.
It suggests that up to $O(1/d^2)$, $\beta_E(d)$ is given
by a random link approximation, leading to a
conjecture even stronger than (\ref{eq_dimscal}).

\section{The random link TSP }
\label{sect_RL}

\subsection{Correspondence with the Euclidean TSP}
\label{sect_RL_model}

Let us now consider a problem at first sight dramatically different
from the Euclidean TSP.
Instead of taking the positions of the $N$ cities to be independent
random variables,
take the lengths $l_{ij}=l_{ji}$ between cities $i$ and $j$
($1\le i,j\le N$) to be independent random variables, identically
distributed according to some $\rho(l)$.  We speak of lengths rather
than distances, as there is no distance metric here.  This
problem, introduced by physicists in the 1980s
\cite{KirkpatrickToulouse,VannimenusMezard} in search of an analytically
tractable form of the traveling salesman problem, is called the
{\it random link TSP\/}.

The connection between this TSP and the Euclidean TSP is not obvious,
as we now have random links rather than random points.  Nevertheless,
one can relate the two problems.  To see this, consider the
probability distribution for the distance $l$ between a fixed pair of
cities ($i$,$j$) in the Euclidean TSP.  This distribution, in the unit
hypercube with periodic boundary conditions, is given for $0\le
l\le 1/2$ by the expression in (\ref{eq_probdens}):
\begin{equation}
\rho(l) = \frac{d\,\pi^{d/2}}{\Gamma(d/2+1)}\,l^{d-1}\mbox{.}
\label{eq_distribution}
\end{equation}

Of course, in the Euclidean TSP the link lengths are by no means
{\it independent\/} random variables: correlations such as the
triangle inequality are present.
However, as noted by M\'ezard and Parisi \cite{MezardParisi_86b},
correlations appear exclusively when considering three or more
distances, since any two Euclidean distances are necessarily
independent.  Let us adopt (\ref{eq_distribution})
as the $l_{ij}$ distribution in the limit of small $l$ for
the random link TSP, where $d$ in this case no longer represents
physical dimension but is simply a parameter of the model.  The
Euclidean and random link problems then have the same small $l$
one- and two-link distributions.  In the large $N$ limit the random
link TSP may therefore be considered, rather than as a separate
problem, as a {\it random link approximation\/} to the Euclidean TSP.
Only joint distributions of three or more links differ between
these two TSPs.  If indeed the correlations involved
are not too important, then the random link $\beta_{RL}(d)$ can be
taken as a good estimate of $\beta_E(d)$.  We shall see that this
is true, particularly for large $d$.

\subsection{Scaling at large N}
\label{sect_RL_N_scaling}

As in the Euclidean case, we are interested in understanding the
$N\to\infty$ scaling law in the random link TSP.  It is relatively simple
to see, following an argument similar to the one in
Section \ref{sect_E_Methodology}, that the nearest neighbor
distances $D_k$ have a probability distribution with a scaling
factor $N^{-1/d}$ at large $N$.  Vannimenus and M\'ezard
\cite{VannimenusMezard} have suggested that the random link 
optimum tour length with $N$ links will then scale as $N^{1-1/d}$,
and the tour will be self-averaging, i.e.,
\begin{equation}
\label{eq_betaRL}
\lim_{N\to\infty} \frac{L_{RL}}{N^{1-1/d}} = \beta_{RL}(d)\mbox{,}
\end{equation}
parallel to the BHH theorem (\ref{eq_bhh}) for the Euclidean case.
This involves the implicit assumption that optimum
tours sample a representative part of the
$D_k$ distribution, so no further $N$ scaling effects are introduced.
The assumption seems reasonable based on the analogy
with the Euclidean TSP, and for our purposes we shall accept here
that $\beta_{RL}(d)$ exists.  However, there is to our knowledge
no mathematical proof of self-averaging in the random link TSP.

Following the discussion of Section \ref{sect_E_Scaling},
let us consider some bounds on the ensemble average
$\langle L_{RL}\rangle$ as derived in \cite{VannimenusMezard}. 
As before, we get a lower bound on $\beta_{RL}(d)$ using
nearest and next nearest neighbor distances.
For an upper bound, the ``strip'' algorithm used in the Euclidean
case (Section \ref{sect_E_Scaling}) cannot be applied to
the random link case.  On the other hand, Vannimenus and M\'ezard
make use of an algorithm called ``greedy'' \cite{PapadimitriouSteiglitz}:
this constructs a non-optimal tour by starting at an arbitrary city,
and then successively picking the link to the nearest available
city until all cities are used once and a closed tour is formed.
At $d>1$, greedy gives rise to tour lengths that are
self-averaging, and leads to the upper bound
\cite{VannimenusMezard}
\begin{equation}
\label{eq_bound}
\beta_{RL}(d)\le
\frac{1}{\sqrt{\pi}}\,\frac{\Gamma(d/2+1)^{1/d}\,\Gamma(1/d)}{d-1}\mbox{.}
\end{equation}
At $d=1$, the presumed scaling (\ref{eq_betaRL}) suggests
that $\langle L_{RL}\rangle$ is independent of $N$, whereas greedy
generates tour lengths which grow as $\ln N$.
There is numerical evidence \cite{Krauth_Thesis,KrauthMezard}, however, 
that the $d=1$ model does indeed satisfy (\ref{eq_betaRL}), and that
$\beta_{RL}(1) \approx 1.0208$.

\subsection{Solution via the cavity equations}
\label{sect_RL_cavity}

Since the work of Vannimenus and M\'ezard,
several groups \cite{MezardParisi_85,Orland,BaskaranFuAnderson_TSP}
have tried to ``solve'' the statistical mechanical problem of the 
random link TSP at finite temperature using the replica method,
a technique developed for analyzing disordered systems
such as spin glasses \cite{MezardParisiVirasoro}.
To date, it has only been possible to obtain part of the
high temperature series of this system \cite{MezardParisi_85}.
In view of the intractability of these replica approaches,
M\'ezard and Parisi have derived an analytical solution using
another technique from spin glass theory, the ``cavity method.''
The details of this approach are beyond the scope of this paper,
and are discussed in several technical articles
\cite{MezardParisi_86b,MezardParisiVirasoro_86,MezardParisiVirasoro}.
For readers acquainted with the language
of disordered systems, however, the broad outline is as follows:
one begins with a representation of the TSP in terms of a Heisenberg
(multi-dimensional spin) model in the limit where the spin
dimension goes to zero.  Under the assumption that this system has only
one equilibrium state (no replica symmetry breaking), M\'ezard
and Parisi have then written
a recursion equation for the system when a new $(N+1)$th spin is
added.  The cavity method then supposes that this new
spin's effect on the $N$ other spins is negligible in
the large $N$ limit, and that its magnetization may be
expressed in terms of the magnetizations of the other spins.

Using this method, Krauth and M\'ezard have derived a
self-consistent equation for the random link TSP, at $N\to\infty$
\cite{KrauthMezard}.  They have determined the
probability distribution of link lengths in the optimum tour in terms
of ${\cal G}_d(x)$, where ${\cal G}_d(x)$ is the solution to the
integral equation
\begin{equation}
{\cal G}_d(x) = \int_{-x}^{+\infty}\frac{(x+y)^{d-1}}{\Gamma(d)}
\left[ 1+{\cal G}_d(y)\right] e^{-{\cal G}_d(y)}\,dy\mbox{.}
\label{caveqn1}
\end{equation}
Their probability distribution leads to the prediction
\begin{equation}
\beta_{RL}(d) =\frac{d}{2\sqrt{\pi}}
\left[ \frac{\Gamma(d/2+1)}
{\Gamma(d+1)}\right]^{1/d}
\int_{-\infty}^{+\infty}{\cal G}_d(x)
\left[ 1+{\cal G}_d(x)\right] e^{-{\cal G}_d(x)}\,dx\mbox{.}
\label{caveqn2}
\end{equation}
These equations can be solved numerically, as well as analytically
in terms of a $1/d$ power series (see next section).  
At $d=1$, Krauth and M\'ezard compared their prediction with
the results of a direct simulation of the random link model; their
numerical study \cite{Krauth_Thesis,KrauthMezard} strongly suggests
that the cavity prediction is exact in this case.  It has been argued,
furthermore, that the cavity method is exact at $N\to\infty$ for
{\it any\/} distribution of the independent random links
\cite{MezardParisiVirasoro}.  Good numerical evidence has been
found for this, notably in the case of the matching problem, a
related combinatorial optimization problem \cite{BKMP}.
The validity of the cavity assumptions therefore does not
appear to be sensitive to the dimension $d$, and we shall assume
that (\ref{caveqn2}) holds for the random link TSP at all $d$.

Krauth and M\'ezard computed the $d=1$ and $d=2$ cases to give
$\beta_{RL}(1) = 1.0208$ and $\beta_{RL}(2) = 0.7251$.
Since $\beta_{RL}(d)$ is taken to approximate $\beta_E(d)$, let
us compare these values with their Euclidean counterparts.
At $d=1$, the Euclidean TSP with periodic boundary
conditions is trivial ($\beta_E(1)=1$); the random link
TSP thus has a 2.1\% relative
excess.  At $d=2$, comparing with $\beta_E(2)=0.7120$ found in
Section \ref{sect_E_Finite}, the random link TSP has
a 1.8\% excess.  In low dimensions, the random link 
results are then a good approximation of
the Euclidean results.  The approximation is better than
Krauth and M\'ezard believed, since they made the comparison
at $d=2$ using the considerably overestimated Euclidean value
of $\beta_E(2)\approx 0.749$ from \cite{BHH}.

Extending the numerical solutions to higher dimensions, at $d=3$ we
find $\beta_{RL}(3) = 0.7100$, which compared with $\beta_E(3)=
0.6979$, has an excess of 1.7\%.
Some further random link values are $\beta_{RL}(4)=0.7322$ and
$\beta_{RL}(5)=0.7639$.
The value at $d=4$ may be compared with the Euclidean estimate
of Johnson {\it et al.\/}\ \cite{Johnson_HK}, $\beta_E(4)\approx
0.7234$, giving
an excess of 1.2\%.  The $\beta_E(d)$ data at $d=1,2,3,4$ therefore
suggest that the random link approximation improves with
increasing dimension.  This leads us to study the limit when $d$
becomes large.

\subsection{Dimensional dependence}
\label{sect_RL_largedimension}

The large $d$ limit was considered by Vannimenus and M\'ezard
\cite{VannimenusMezard}.
For $\beta_{RL}(d)$, the lower bound obtained from $\langle
D_1(N,d)+D_2(N,d)\rangle /2$ by way of (\ref{eq_nn}) and the upper
bound given in (\ref{eq_bound}) differ at large $d$ only
by $O(1/d)$, giving:
\begin{equation}
\label{eq_RL_dimscal}
\beta_{RL}(d)=\sqrt{\frac{d}{2\pi e}}\,(\pi d)^{1/2d}
\left[ 1 + O\left( \frac{1}{d}\right) \right]\mbox{.}
\end{equation}
Note that this {\it exact\/} result is the random link analogue of the
Euclidean conjecture (\ref{eq_dimscal}).

\begin{figure}[!b]
\begin{center}
\vspace{20pt}
\begin{picture}(282,271)
\epsfig{file=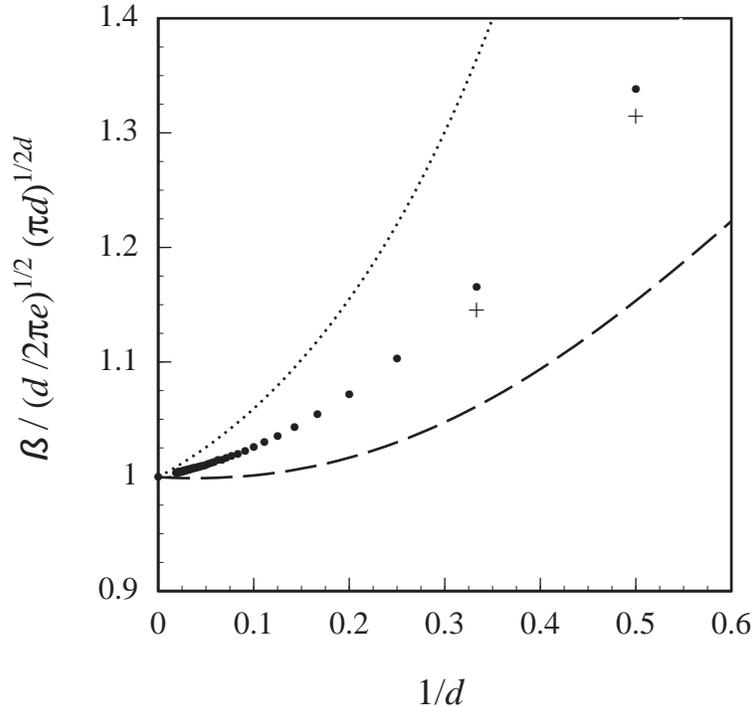}
\end{picture}
\caption{\small Dimensional dependence of rescaled random link TSP optimum,
shown by small points, between converging ``greedy'' upper bound (dotted
line) and nearest-neighbors lower bound (dashed line).  Plus signs
at $d=2$ and $d=3$ show Euclidean results for comparison.}
\label{fig_rl}
\vspace{-20pt}
\end{center}
\end{figure}

For values of $d\le 50$, we have calculated $\beta_{RL}(d)$ numerically
using the cavity equations (\ref{caveqn1}) and (\ref{caveqn2}).  The
results are shown in Figure \ref{fig_rl}, along with the converging
upper and lower bounds, and our low $d$ Euclidean results.

\def\hatG{{\tilde {\cal G}}}

For large $d$, we may see whether the cavity equations are compatible
with (\ref{eq_RL_dimscal}) by solving them analytically in
terms of a $1/d$ power series.  Define $\hatG_d(x)\equiv
{\cal G}_d\bigl(\Gamma(d+1)^{1/d}\,[1/2 + x/d]\bigr)$.
(\ref{caveqn1}) may then be written:
\begin{eqnarray}
\hatG_d(x)&=&\int_{-x-d}^{+\infty}\left( 1+\frac{x+y}{d}\right)^{d-1}
\left[ 1+\hatG_d(y)\right] e^{-\hatG_d(y)}\,dy\\
&=&\int_{-x-d}^{+\infty} e^{x+y}\left[ 1-\frac{1}{d}\left( x+y+
\frac{(x+y)^2}{2}\right) + O\left(\frac{1}{d^2}\right)\right]
\left[ 1+\hatG_d(y)\right] e^{-\hatG_d(y)}\,dy\mbox{.}
\end{eqnarray}
Strictly speaking, the expansion of $(1+[x+y]/d)^{d-1}$ is only
valid in the interval $-x-d<y<-x+d$; however, for large $y$ it
can be shown that $\hatG_d(y)\sim y^d$, so the $e^{-\hatG_d(y)}$
term in the integrand makes the $y>-x+d$ contribution exponentially
small in $d$.

Furthermore, extending the integral's lower limit to include the
region $y<-x-d$ also
contributes a remainder term exponentially small in $d$.  If we
write the integral with its lower limit at $y=-\infty$, the equation
may be solved:
\begin{equation}
\hatG_d(x)=\sqrt{2}e^x\left[ 1 - \frac{1}{d}\left(\frac{x^2}{2}+
x\,\frac{3-\ln 2 -2\gamma}{2}-\frac{(\ln 2 +2\gamma)^2+6\ln 2
+12\gamma -9}{8}\right)+O\left(\frac{1}{d^2}\right)\right]\mbox{,}
\end{equation}
where $\gamma$, we recall, represents Euler's constant.
Using (\ref{caveqn2}), we
then find
\begin{equation}
\label{cavresult}
\beta_{RL}(d) = \sqrt{\frac{d}{2\pi e}}\,(\pi d)^{1/2d}
\left[ 1+\frac{2-\ln 2-2\gamma}{d} +
O\left(\frac{1}{d^2}\right)\right]\mbox{,}
\end{equation}
which is perfectly compatible with (\ref{eq_RL_dimscal}).  This provides
further evidence that the cavity method is exact for the random link
TSP.

\subsection{Renormalized random link model at large $d$ }
\label{sect_RL_renorm}

We can motivate the large $d$ scaling found in the previous
section by examining a different sort of random link TSP.
Consider a new ``renormalized'' model where link ``lengths''
$x_{ij}$ are obtained
from the original $l_{ij}$ by the linear transformation
$x_{ij}\equiv d [l_{ij} -\langle D_1(N,d)\rangle ] /
\langle D_1(N,d)\rangle$.
Note that the $x_{ij}$ may take on negative values, and that
the nearest neighbor length in this new model has mean zero.
Since the transformation is linear, there is a direct
equivalence between the renormalized $x_{ij}$ and original
$l_{ij}$ TSPs, and the two have the same optimum tours.
The renormalized optimum tour length $L_x$ may then be given in terms
of the original tour length $L_l$ by
\begin{equation}
L_x = d\,\frac{L_l-N\langle D_1(N,d)\rangle}{\langle D_1(N,d)\rangle}\mbox{.}
\label{eq_renorm}
\end{equation}

Now take $N \to \infty$ and $d \to \infty$.  It may
be seen from the $l_{ij}$ distribution (\ref{eq_distribution})
and the $\langle D_1(N,d)\rangle$ expansion (\ref{eq_largek})
that the random variables $x_{ij}$ have the $d$-independent
probability distribution $\rho(x)\sim N^{-1} \exp{(x - \gamma)}$.
Also, in the large $N$ limit, since $L_l$ scales as $N^{1-1/d}$ and
$\langle D_1\rangle$ scales as $N^{-1/d}$, we expect
$\langle L_x\rangle\sim N\mu$ for some $\mu$ which must be,
like $\rho(x)$, independent of $d$.
Then, from (\ref{eq_renorm}), the TSP in the original $l_{ij}$
variables satisfies
\begin{equation}
\langle L_l\rangle\sim N\langle D_1(N,d)\rangle\left[ 1 + \frac{\mu}{d}
+ O\left(\frac{1}{d^2}\right)\right]\mbox{,}
\end{equation}
or, using the expansion (\ref{eq_largek}),
\begin{equation}
\beta_{RL}(d)=\sqrt{\frac{d}{2\pi e}}\,(\pi d)^{1/2d}
\left[ 1 + \frac{\mu -\gamma}{d} +
O\left(\frac{1}{d^2}\right)\right]\mbox{.}
\end{equation}
This result may be compared with our cavity solution of (\ref{cavresult}),
where the $1/d$ coefficient is equal to $2-\ln 2 -2\gamma$.  If the
cavity method is correct at $O(1/d)$, which we strongly believe
is the case, then a direct solution of the renormalized model should
give $\mu=2-\ln 2 -\gamma$.  Work is currently in progress to test
this claim by numerical methods.

\subsection{Large $d$ accuracy of the random link approximation }
\label{sect_RL_estimate}

Since the random link model is considered to be an approximation
to the Euclidean case, it is natural to ask whether the
approximation becomes exact as $d \to \infty$.  In this
section we argue that: (i) in stochastic TSPs, 
good tours can be obtained using almost exclusively low order neighbors; 
(ii) the geometry inherent in the Euclidean TSP leads to
$\beta_E(d) \le \beta_{RL}(d)$ in all dimensions $d$;
(iii) the relative error of the random link approximation
decreases as $1/d^2$ at large $d$.
All three claims are based on a theoretical analysis of the Lin-Kernighan
(LK) heuristic algorithm for constructing near-optimal tours.

\begin{figure}[t]
\begin{center}
\begin{picture}(231,224)
\epsfig{file=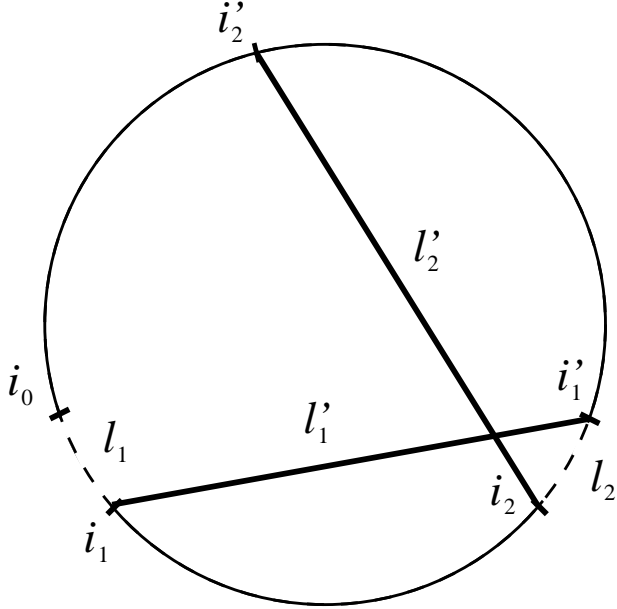}
\end{picture}
\caption{\small Recursive construction of removed links (dashed lines) and
added links (bold lines) in an LK search.}
\label{fig_LK}
\end{center}
\end{figure}

The LK algorithm works as follows \cite{LinKernighan,JohnsonMcGeoch}.
An {\it LK search\/} starts with
an arbitrary tour.  The principle of the search is to substitute links
in the tour recursively, as illustrated schematically in
Figure \ref{fig_LK}.  The first
step consists of choosing an arbitrary starting city $i_0$.  Call $i_1$
the next city on the tour, and $l_1$ the link between the two.  Now
remove this link.  Let $i_1'$ be the nearest neighbor to $i_1$ that
was not connected to $i_1$ on the original tour, and let $l_1'$ be
a new link connecting $i_1$ to $i_1'$.  We now have not a tour but a
``tadpole graph,'' containing a loop with a tail attached to it at $i_1'$.
At this point, call $i_2$ one of the cities next to $i_1'$ on the original
tour, and remove the link $l_2$ between the two.  There are two possibilities
for $i_2$ (and thus $l_2$): LK chooses the one which, if we were to put in
a new link between $i_2$ and $i_0$, would give a single closed tour.
Now as before, let $i_2'$ be the nearest neighbor of $i_2$ that was
not connected to $i_2$ on the original tour, and let $l_2'$ be a new
link between the two.  This gives a new tadpole.  The process continues
recursively in this manner, with the vertex hopping around while the
end point stays fixed, until no new tadpoles are found.
At each step, LK chooses the new $i_m$ so as to allow the path to
be closed up between $i_m$ and $i_0$, forming a single tour; the
result of the LK search is then the best of all such closed up tours.
The {\it LK algorithm\/} consists of repeating these LK searches
on different starting points $i_0$, each time using the current best
tour as a starting tour, until no further tour improvements are
possible.

Let us first sketch why the LK algorithm leads to tours which use
only links between ``near'' neighbors, where ``near'' means
that the neighborhood order $k$ is small and does not grow with $d$. 
Consider any tour where a significant fraction of the links
connect distant neighbors (large $k$).  The links $l_m'$ which
the LK search substitutes for the $l_m$ are, by definition, between very
near neighbors ($k\le 3$).  As long as many long links exist, the
probability at each step of substituting a near neighbor in place
of a far neighbor is significant.  Towards the beginning of an LK
search this probability is relatively constant, so
the {\it expected\/} tadpole length
will decrease linearly with the number of steps.
Even taking into account the fact that closing up the path between
$i_m$ and $i_0$ might require inserting a link with $k>3$, there is a high
probability as $N\to\infty$ that the improvement in tadpole length far
outweighs this cost of closing the tour.  Thus for stochastic TSPs,
regardless of $d$, the LK algorithm can at large $N$ replace all
but a tiny fraction of the long links with short links.  
It follows that
in accordance with our Euclidean TSP assumption of Section
\ref{sect_E_Large_d}, the ``typical'' $k$ used in the optimum tour
remains small at large $d$.  This provides very powerful support for the
$\beta_E(d)$ conjectures (\ref{eq_dimscal}) and (\ref{eq_Bertsimas}).
A consequence, making use of the exact asymptotic
$\beta_{RL}(d)$ result (\ref{eq_RL_dimscal}), is that the relative difference
between $\beta_E(d)$ and $\beta_{RL}(d)$ is at most of $O(1/d)$.

Our second argument concerns why $\beta_{RL}(d)$ must be greater than
$\beta_E(d)$ at all $d$.  For the random link TSP there is no
triangle inequality, which means that given two edges of a triangle,
the third edge is on average longer than it would be for the Euclidean
TSP.  Applying this to our LK search, we can expect the link between
$i_m$ and $i_0$ closing up the tour to be
longer in the random link case than in the
Euclidean case.  Thus on average, the LK algorithm will find longer
random link tours than Euclidean tours.  In fact, this property holds
as well for any LK-like algorithm where the method of choosing the
$l_m$ and $l_m'$ links is generalized.  If the algorithm were to allow
{\it all\/} possibilities for $l_m$ and $l_m'$, we would be sure of
obtaining the exact optimum tour, given a long enough search.
In that case, the inequality on
the tour lengths found by our algorithm leads directly to
$\beta_{RL}(d)>\beta_E(d)$.
Not surprisingly, the numerical data confirm this inequality at
$d$ up to 4 (although one should be cautious when applying the argument
at $d=1$).  Note also that the inequality in itself implies 
conjectures (\ref{eq_dimscal}) and (\ref{eq_Bertsimas}) for the
Euclidean model, since it supplies precisely the upper bound we
need on $\beta_E(d)$.

Finally let us explain why the relative difference between $\beta_{RL}(d)$
and $\beta_E(d)$ should be of $O(1/d^2)$.  This involves quantifying
the tour length improvement discussed above.  It is clear that any
non-optimal tour can be improved to the point where links
are mostly between neighbors of low order.  If LK, or a generalized
LK-like algorithm, is able to improve the tour further, the relative
difference in length
will be of $O(1/d)$; we see this from (\ref{eq_largek}),
noting that the neighborhood order $k$ is small both before and after
the LK search.
Now we need to quantify the {\it probability\/} that LK indeed succeeds in
improving the tour.  We may consider the vertex of the LK tadpole graph
as executing a random walk, in which case the probability of closing up
a tour by a sufficiently short link is equivalent to the probability
of the random walk's end-to-end distance being sufficiently small.
In that case it may be shown that, over the course of an LK search,
the probability of successfully closing a random link tour minus
the probability of successfully closing a Euclidean tour scales
at large $d$ as $2/(d-2)$.  From this, we conclude that
improvements in the Euclidean model are $O(1/d)$ more probable
than in the random link model.  Now, the relative tour length
improvement for the Euclidean TSP compared to the random link TSP
is simply the relative tour length improvement {\it when a better
tour is found\/}, times the probability of finding a better tour ---
hence $O(1/d^2)$.  If we consider a generalized LK search as described
in the previous paragraph, where the algorithm necessarily finds the
true optimum, then this result applies to the exact $\beta$s: the
relative difference between $\beta_{RL}(d)$ and $\beta_E(d)$ will
scale at large $d$ as $1/d^2$.

Three comments are in order concerning this surprisingly good
accuracy of the random link approximation.  First, the
factor $2/(d-2)$ is only appropriate for large $d$.  It is not
small even for $d=4$.  (Its divergence at $d=2$ is associated
with the fact that a two-dimensional random walk returns to its
origin with probability 1.)  We therefore expect the $1/d^2$
scaling to become apparent only for $d\ge 5$, beyond the range
of our numerical data.  Second, we have seen that the coefficient
of the $1/d$ term in $\beta_{RL}(d)$ may be obtained by the cavity
method.  Assuming that this method is correct and that
$\beta_{RL}(d)$ and $\beta_E(d)$ do indeed converge as $1/d^2$,
this leads to a particularly strong conjecture for the Euclidean TSP:
\begin{equation}
\label{eq_conjecture_text}
\beta_E(d) = \sqrt{\frac{d}{2\pi e}}\,(\pi d)^{1/2d} \left[
1+\frac{2-\ln 2-2\gamma}{d} + O\left(\frac{1}{d^2}\right)\right]\mbox{.}
\end{equation}
Third, this type of LK analysis can in fact be extended to many other
combinatorial optimization problems, such as the assignment, matching
and bipartite matching problems.  In these cases, we expect the
random link approximation
to give rise to a $O(1/d^2)$ relative error just as in the TSP.

\section{Summary and conclusions}
\label{sect_conclusions}

The first goal in our work has been to investigate the
finite size scaling of $L_E$, the optimum Euclidean traveling
salesman tour length, and to obtain precise estimates
for its large $N$ behavior.  Motivated by a remarkable universality
in the $k$th nearest neighbor distribution, we have found
that under periodic boundary conditions, the convergence of
$\langle L_E\rangle /N^{1-1/d}$ to its limit $\beta_E(d)$ is
described by a series in $1/N$.  This has enabled us to extract
$\beta_E(2)$ and $\beta_E(3)$ using numerical simulations at small
values of $N$, where errors are easy to control.  Furthermore,
thanks to a bias-free variance reduction method (see
\ref{app_statistical}), these estimates are extremely precise.

Our second goal has been to examine the random link TSP, where there
are no correlations between link lengths.  We have considered it
as an approximation to the Euclidean TSP, in order to understand
better the dimensional scaling of $\beta_E(d)$.  For small $d$, we
have used the cavity method to obtain numerical values of the random
link $\beta_{RL}(d)$.  Comparing these with our numerical values for
$\beta_E(d)$ shows that the random link approximation is remarkably
good, accurate to within 2\% at low dimension.  For large $d$,
we have solved the cavity equations analytically to give
$\beta_{RL}(d)$ in terms of a $1/d$ series.
We have then argued, using a theoretical analysis of iterative tour
improvement algorithms, that the relative difference between $\beta_{RL}(d)$
and $\beta_E(d)$ decreases as $1/d^2$.  This leads to our conjecture
(\ref{eq_conjecture_text})
on the large $d$ behavior of $\beta_E(d)$, specifying both its
asymptotic form and its leading order correction.

Let us conclude with some remaining open questions.  First of all,
while the cavity method most likely gives the exact result for
the random link TSP, we would be interested in seeing this argued
on a more fundamental physical level.  Readers with a background in
disordered systems will recognize that the underlying assumption of
a unique equilibrium state is false in many NP-complete problems,
and in particular in the spin-glass problem that has inspired the
cavity method.  What makes the TSP different?
Second of all, our renormalized random link model provides an
alternate approach to finding the $1/d$ coefficient of the power
series in $\beta_{RL}(d)$, and could prove a useful test of the
cavity method's validity.  A solution to the renormalized model
using heuristic methods appears within reach.
Third of all, the $O(1/d^2)$ convergence of the random link approximation
merits further study, from both numerical and analytical perspectives.
Numerically, Euclidean simulations at $d\ge 5$ could
provide powerful support for the form of the convergence, and thus
for our conjecture (\ref{eq_conjecture_text}).  Analytically,
the qualitative arguments presented in Section \ref{sect_RL_estimate},
based on the LK algorithm, could perhaps be refined
by a more quantitative approach.  Lastly, it is worth noting
that the $O(1/d^2)$ convergence should apply equally well to the
{\it distribution\/} of link lengths in the optimum tour.  The
random link prediction for this distribution can be
obtained from the cavity method \cite{KrauthMezard}; an interesting
test would then be to compare it with simulation results for
the true Euclidean distribution.

\section*{Acknowledgments}
\label{sect_ack}
We thank E.~Bogomolny, D.S.~Johnson, M.~M\'ezard, S.W.~Otto and
N.~Sourlas for fruitful discussions.  NJC acknowledges a grant from
the Human Capital and Mobility Program of the Commission of the
European Communities; JBdM acknowledges financial support from the
French Ministry of Higher Education and Research; OCM acknowledges
support from NATO travel grant CRG 920831 and from the Institut
Universitaire de France.  

\vspace{1 cm}

\setcounter{subsection}{0}
\renewcommand{\thesubsection}{Appendix \Alph{subsection}}
\renewcommand{\theequation}{\Alph{subsection}.\arabic{equation}}
\newpage

\subsection{ }
\subsection*{Overview of the numerical methodology}
\label{app_methodology}
\setcounter{equation}{0}

In the following, we discuss the procedures used to obtain the raw
data from which $\beta_E(d)$ and the finite size scaling coefficients
are extracted.  Two major problems must be solved
in order to get good estimates of $\beta_E(N,d)$.  First, 
$\beta_E(N,d)$ is defined as an ensemble average
$\langle L_E(N,d)\rangle / N^{1-1/d}$, but is measured by a numerical
average over a finite sample of instances.  The instance-to-instance
fluctuations in $L_E$ give rise to a statistical error, which decreases
only as the inverse square root of the sample size.  Keeping the
statistical error down to acceptable levels could require inordinate
amounts of computing time.  We therefore find it useful to introduce
a variance reduction 
trick: instead of measuring $L_E$, we measure $L_E - \lambda L^{*}$,
where $\lambda$ is a free parameter and $L^{*}$ can be any quantity
which is strongly correlated with $L_E$. 
Details are given in \ref{app_statistical}.

A second and more basic problem is that it is computationally costly
to determine the optimal tour lengths for a large number of instances,
precisely because the TSP is an NP-complete problem.  The most sophisticated
``branch and cut'' algorithms can take minutes on a workstation to
solve a single instance of size $N\le 100$ to optimality.
However, we do not need to guarantee optimality: the statistical
error in $\beta_E(N,d)$ already limits the quality of our
estimate, and so an additional (systematic) error in $L_E$ is
admissible as long as it is negligible compared to the statistical
error.  We may thus use fast heuristics to measure $L_E$, rather
than exact but slower algorithms.  This is discussed further
in \ref{app_systematic}.

\subsection{ }
\subsection*{Statistical errors and a variance reduction trick}
\label{app_statistical}
\setcounter{equation}{0}

Consider estimating $\langle L_E(N,d)\rangle$ at a given $N$ by
sampling over many instances.  If we have $M$ independent
instances, the simplest estimator for $\langle L_E(N,d)\rangle$ 
is $\overline{L_E(N,d)}$, the numerical average over the $M$
instances of the minimum tour lengths.  This estimator
has an expected statistical error $\sigma(M) = \sigma_{L_E} / \sqrt{M}$,
where $\sigma_{L_E}$ is the instance-to-instance standard
deviation of $L_E$.

Now let us define $L_k$ to be the sum, over all cities, of $k$th
nearest neighbor distances.  $\langle L_k\rangle$ is its ensemble
average; in terms of the notation used earlier in the text,
$\langle L_k\rangle=N\langle D_k\rangle$.
It has been noted by Sourlas \cite{Sourlas} that $L_E$ is strongly
correlated with $L_1$, $L_2$ and $L_3$.  He therefore suggested
reducing the statistical error in $\langle L_E\rangle$ using the estimator
\begin{equation}
E_S = \langle L_{123}\rangle\,\overline{L_E/L_{123}}\mbox{,}
\end{equation}
where $L_{123}$ is the arithmetic mean of $L_1$, $L_2$ and $L_3$.
The ensemble average $\langle L_{123}\rangle$ can be calculated
analytically from (\ref{eq_nn}), and
so the variance of $E_S$ comes from fluctuations in the ratio
$L_E/L_{123}$.  If $L_E$ were a constant factor times $L_{123}$,
this estimator would of course be perfect, i.e., it would have
zero variance.  This is not the case, however, and furthermore
the use of a ratio biases the Sourlas estimator: its true mathematical
expectation value differs from $\langle L_E(N,d)\rangle$ by
$O(1/N)$.
To improve upon this, we have introduced our own bias-free estimator
\cite{MartinPercus_EJOR}:
\begin{equation}
E_{M\mbox{-}P} = \lambda\langle L_{12}\rangle + \overline{L_E -
\lambda L_{12}}\mbox{,}
\end{equation}
where $L_{12}$ is the arithmetic mean of $L_1$ and $L_2$,
and $\lambda$ is a free parameter.
Our estimator
has a reduced variance because $L_E$ and $L_{12}$ are correlated.
It is easy to show that the variance of $E_{M\mbox{-}P}$ is
minimized at a unique value of $\lambda$,
$\lambda^* = C(L_E,L_{12})\,\sigma_{L_E}/\sigma_{L_{12}}$,
where $C(A,B)\equiv\bigl\langle\, (A - \langle A\rangle )\,
(B - \langle B\rangle )\,\bigr\rangle /\sigma_A \sigma_B$ is the
correlation coefficient of $A$ and $B$.
The variance then becomes
$\sigma_{L_{M\mbox{-}P}}^2 = \sigma_{L_E}^2 [1 - C^2(L_E,L_{12})] / M$.
Empirically, we have found this variance reduction procedure to be
quite effective, since $\sqrt{1-C^2}\approx 0.38$ at $d=2$ and
$\sqrt{1-C^2}\approx 0.31$ at $d=3$.  The statistical error is thus
reduced by about a factor of 3; this means that for a given error,
computing time is reduced by about a factor of 10.

\subsection{ }
\subsection*{Control of systematic errors}
\label{app_systematic}
\setcounter{equation}{0}

Our procedure for estimating $L_E$ at a given instance involves
running a good heuristic $m$ times from random starts on that
instance, and taking the best tour length found in those $m$ trials. 
The expected systematic error can be found from the frequencies
with which each local optimum appears in a large number of test
trials.  (This large number must be much greater than $m$, the
actual number of trials used in production runs.)
The measurement is performed on a sufficiently large sample of instances,
from which we extract the {\it average\/} size of the systematic
error in $\langle L_E(N,d)\rangle$
as a function of $m$.  We have found that in practice, this
error is dominated by those infrequent instances where a
sub-optimal tour is obtained with the highest frequency.

As $N$ increases, the probability of not finding the true optimum
increases rather fast; for a given heuristic, it is thus necessary
to increase $m$ with $N$ in such a way that the systematic error
remains much smaller than the statistical error.
If the heuristic is not powerful enough, $m$ will be too
large for the computational resources. For our purposes,
we have found that
the Lin-Kernighan heuristic \cite{LinKernighan}
is powerful enough for the smaller values
of $N$ ($N \le 17$). For $20 \le N \le 100$, it was more efficient to
switch to Chained Local Optimization
(CLO) \cite{MartinOtto_AOR,MartinOttoFelten_CS}, 
a more powerful heuristic
which can be thought of as a generalization of simulated annealing.
(When the temperature parameter is set to zero so that no up-hill moves
are accepted, as was the case for our runs, CLO with embedded
Lin-Kernighan is called ``Iterated Lin-Kernighan''
\cite{Johnson_ILK,MartinOttoFelten_ORL}.)
With these choices, using in two dimensions $m=10$ for $N\le 17$ (LK),
$m=5$ for $N=30$ and $m=20$ for $N=100$ (CLO), we have kept systematic
errors to under 10\% of the statistical errors.

\subsection{ }
\subsection*{Bounding $\beta_{E}(d)$ using the bipartite matching problem}
\label{app_heuristics}
\setcounter{equation}{0}

Given two sets of $N$ points $P_1,\dots,P_N$ and $Q_1,\dots,Q_N$ in
$d$-dimensional Euclidean space, the
bipartite matching (BM) problem asks for the minimum matching
cost $L_{BM}$
between the $P_i$s and the $Q_i$s, with the constraint that only links
of the form $P-Q$ are allowed.  The cost of a matching is equal to the
sum of the distances between matched pairs of points.
When points $P_i$ and $Q_i$ are chosen at random
in a $d$-dimensional unit hypercube,
it is natural to expect $L_{BM}/N^{1-1/d}$
to be self-averaging as $N \to \infty$. 
To date, a proof of this property has not been given, 
even though the self-averaging of the analogous quantity in the
more general matching problem (where links $P-P$ and
$Q-Q$ are allowed as well) can be shown at all $d$ in essentially
the same way as for the TSP,
following arguments developed by Steele \cite{Steele}.
For $d=1$, it is in fact known that self-averaging 
{\it fails\/} in the BM.
For large $d$, however, let us assume that $L_{BM}/N^{1-1/d}$ 
does converge to some $\beta_{BM}(d)$ in the large $N$ limit.

We shall now derive a bound for the Euclidean TSP constant $\beta_{E}(d)$
in terms of $\beta_{BM}(d)$.
Consider $K$ disjoint sets $S_1,\dots,S_K$, together forming a large
set $S\equiv S_1\cup\cdots\cup S_K$, and let each $S_i$ contain
$N$ random points in the $d$-dimensional unit hypercube.
Construct the $K$ minimum matchings $S_1-S_2, S_2-S_3,\dots,S_{K-1}-S_K$
and $S_K-S_1$.  Starting at any point in $S_1$, generate a loop (a
closed path) in $S$ by following the
matchings $S_1-S_2$, $S_2-S_3$, $\dots$ until the path returns to its
starting point. The set of all such distinct loops $\Omega_1,\dots,\Omega_M$
($M\le N$) is then equivalent to the set $S$, and
furthermore the sum of the loop lengths is equal to the sum of all
minimum matchings costs $(L_{BM})_{S_i-S_{i+1}}$.  (Note that
$(L_{BM})_{S_K-S_{K+1}}$ is defined as $(L_{BM})_{S_K-S_1}$.)

Now, consider the optimum TSP tour through all the points of $S_1$.
Construct a giant closed path visiting every point in $S$ at least
once, by substituting into this TSP tour the loops
$\Omega_1,\dots,\Omega_M$ in place of their starting points in $S_1$.
Using standard techniques \cite{KarpSteele}, we can construct from
this path of length $(K+1)N$ a shorter closed path of length $K N$ which
visits every point in $S$ exactly once.  For the Euclidean TSP tour
length $L_E$, we then obtain the inequality
\begin{equation}
(L_{E})_S\le (L_{E})_{S_1} + \sum_{i=1}^K (L_{BM})_{S_i-S_{i+1}}\mbox{.}
\end{equation}

If $S$ consists of random points chosen independently and
uniformly in the unit hypercube, then averaging over all configurations,
dividing by $N^{1-1/d}$ and taking the limit $N\to \infty$, we find
\begin{equation}
K^{1-1/d} \beta_{E}(d) \le \beta_{E}(d) + K\beta_{BM}(d)\mbox{.}
\end{equation}
Letting $K=d$, this gives in the large $d$ limit $\beta_{E}(d)\le
\beta_{BM}(d)$.
Based on analogies with other combinatorial optimization problems
\cite{BertsimasVanRyzin}, $\beta_{BM}(d)$ is expected to scale
as $\sqrt{d/2\pi e}$ when $d\to\infty$.  In that case, $\beta_{E}(d)$
too must satisfy the Bertsimas-van Ryzin conjecture (\ref{eq_Bertsimas}).

\end{document}